\documentstyle[12pt,aaspp4]{article}

\def\kms{km s$^{-1}$}

\def\gtsim{\ {\raise-0.5ex\hbox{$\buildrel>\over\sim$}}\ }
\def\ltsim{\ {\raise-0.5ex\hbox{$\buildrel<\over\sim$}}\ }
\slugcomment{Submitted to The Astrophysical Journal}

\begin{document}

\title{MACHO Project Photometry of RR Lyrae Stars in the Sgr Dwarf Galaxy}

\author {C.  Alcock$^{1,2}$, R.A.  Allsman$^{1,4}$, D. R. Alves $^{1,9}$,
T.S. Axelrod$^{1,4}$, A. Becker$^{2,3}$, D.P. Bennett$^{1,2}$, K.H. Cook$^{1,2}$, 
K.C. Freeman$^4$, K. Griest$^{2,5}$, J.A. Guern$^5$, M.J. Lehner$^5$, 
S.L. Marshall$^{1}$, D. Minniti$^{1}$,
B.A. Peterson$^4$, M.R. Pratt$^{3,6}$, P.J. Quinn$^4$, A.W. Rodgers$^4$, C.W.
Stubbs$^{2,3,6}$, W. Sutherland$^7$, and D.L. Welch$^{8}$}
\affil { (The MACHO Collaboration) }

\altaffiltext{1}{Lawrence Livermore National Laboratory, Livermore, CA 94550\\
E-mail:  alcock, alves, robynallsman, tsa, bennett, kcook, dminniti, stuart@llnl.gov}

\altaffiltext{2}{Center for Particle Astrophysics, University of California,
Berkeley, CA 94720}

\altaffiltext{3}{Department of Astronomy, University of Washington,
Seattle, WA 98195 \\
E-mail: stubbs, becker@astro.washington.edu}

\altaffiltext{4}{Mt.  Stromlo and Siding Spring Observatories, Australian Nation
al
University, Weston, ACT 2611, Australia \\ E-mail: kcf, peterson, 
pjq, alex@merlin.anu.edu.au}

\altaffiltext{5}{Department of Physics, University of California,
San Diego, CA 92093 \\
E-mail: griest, jguern, matt@astrophys.ucsd.edu}

\altaffiltext{6}{Department of Physics, University of California,
Santa Barbara, CA 93106 \\
E-mail: mrp@lensing.physics.ucsb.edu}

\altaffiltext{7}{Department of Physics, University of Oxford,
Oxford OX1 3RH, U.K.\\
E-mail: wjs@oxds02.astro.ox.ac.uk}

\altaffiltext{8}{Department of Physics and Astronomy, McMaster University,
Hamilton, ON L8S 4M1, Canada\\
E-mail: welch@physics.mcmaster.ca}

\altaffiltext{9}{Department of Physics, University of California, Davis, CA 95616}

\begin{abstract}
We report the discovery of 30 type a,b RR Lyrae (RRab)
which are likely members of the
Sagittarius (Sgr) dwarf galaxy. Accurate positions, periods, amplitudes and 
magnitudes are presented. 
Their distances are determined with respect to
RRab in the Galactic bulge found also in the MACHO 1993 data. 
For R$_{\odot} = 8$ kpc,
the mean distance to these stars is $D = 22 \pm 1$ kpc,
smaller than previous determinations for this galaxy. This indicates that Sgr 
has an elongated main body extending for more than 10 kpc,
which is inclined along the line of sight,
with its northern part (in Galactic coordinates) closer to us.
The size and shape of Sgr give clues about the past history of this galaxy.
If the shape of Sgr follows the direction of its orbit, 
the observed spatial orientation suggests that
Sgr is moving away from the Galactic plane.
Also, Sgr stars may be the sources of some of the microlensing events
seen towards the bulge.
\end{abstract}

\section{Introduction}

The dwarf galaxy
Sgr was discovered by Ibata et al. (1994) in the direction of the Galactic 
bulge. This is the closest dSph, located at about 25 kpc, and moving
away from us at 160 \kms. The bulk of the stars in this
galaxy is very old, 10$^{10}$ yr, with a range in metallicities,
$[Fe/H] = -0.5$ to $-1.3$ (Mateo et al. 1995, Sarajedini \& Layden 1995,
Fahlman et al. 1996). 
Sgr is close to pericenter, and it is being torn apart by Galactic 
tidal forces, judging from its elongated appearance in the sky (Piatek 
\& Pryor 1995, Velazquez \& White 1995).
Different numerical simulations aimed at tracing the past history 
of Sgr give density maps and velocity fields which agree with the available
observations (Velazquez \& White 1995, Johnston et al. 1995). 
It has not yet been possible, however,
to distinguish among a wide range of orbital parameters and models,
nor to determine the direction of motion of Sgr (i.e., are we observing
it pre- or post-pericenter, and before or after crossing the Galactic plane?).

The MACHO project has identified more than 38000 
variable stars in the bulge
fields from the 1993 data (Cook et al. 1995). Certain variable stars of 
different types (SX Phe, Cepheids,
W UMa, RR Lyrae, Miras), are good distance indicators, and
can be used as probes to study the density of Sgr and of the 
different Galactic components along the line of sight towards the bulge.   
The RR Lyrae stars are by far the best distance 
indicators for an older
population object like the Sgr dwarf galaxy (e.g. Nemec et al. 1994), 
and are the focus of the present study. 
Our goals are three-fold: 
1. identify RRab in Sgr; 
2. determine their distance; and 
3. measure the shape, orientation, and total extension of this galaxy.

RR Lyrae stars belonging to Sgr were first discovered by Mateo et al. (1995) in a field
centered in the galaxy, adjacent to the globular cluster M~54, which is 
also associated with Sgr (Da Costa \& Armandroff 1995).   They determined 
the distance to Sgr, $(m-M)_0 = 17.02 \pm 0.19$ based on 9 RR Lyrae.  More 
recently, Alard (1996) found more than 300 RR Lyrae stars in a field located in
the northern extension of Sgr (in Galactic coordinates), and Mateo et al.
(1996) found 3 RR Lyrae stars in a field next to the globular cluster M~5, 
in the southern extension of Sgr (in Galactic coordinates). 
The 1993 MACHO fields, located to the north of these previous studies
(in Galactic coordinates), add new data, confirming the large extension of Sgr.

\section{The RRab Sample}

The system and data collection of the MACHO experiment are described by 
Alcock et al. (1996). Here we consider only the first year data acquired from
1993 February 27 to September 3. There are about 2300 images of 24 bulge fields,
containing 12.6 million stars total. The photometric measurements are 
made with SoDoPhot, derived from DoPhot (Schechter et al. 1994). 
Light curves for stars identified as variable are
phased in order to find periods (Cook et al. 1995). A typical two-color
light curve has about 100 points. 

The selection of the RRab sample is relatively simple.
The variables in the MACHO 1993 data for
which a period can be identified are plotted in an amplitude--period plane.
RRab have periods within the range 0.3--0.8 days, and amplitudes within
the range 0.1--1.2 mag. However, for periods
smaller than 0.4 days there are larger numbers of RRc variables, with
smaller amplitudes and
typical light curve shapes that make them difficult to discriminate
from contact binaries. Therefore our selection included
periodic variables with amplitudes $>$ 0.2 mag in the MACHO blue band
($V_{Macho}$), and
periods within 0.4-0.7 d. The tighter cut in the amplitudes is imposed
to secure good data. About 1700 stars are selected within those cuts.  
The $V_{Macho}$ and $R_{Macho}$
light curves of all these candidates were inspected in both passbands to decide 
which ones are bonafide RR Lyrae stars (1173 stars in total). 
Of these, 44 are duplicates
because they are found in the overlap region of MACHO fields. 
The other 600 stars are mostly contact binaries. Some period aliases are 
also rejected (see Alcock et al. 1996). 
In summary, we have imposed stringent selection criteria which
ensure the good quality of the
sample. The price paid is that the sample is not complete.

Most of the RRab in the final sample belong to the Galactic bulge. 
Their magnitudes peak at $V \sim 16$, which places them at about 8 kpc.
A few, however, are more than twice as far away (30 stars total).
Not all of them can be halo RRab, according to
current halo models (e.g. Saha 1985).
These distant RRab belong to the Sgr galaxy.

Table 1 lists the MACHO
catalog of RRab's in Sgr, containing 30 entries. We give the
MACHO identification (field and location),
the positions in equatorial and Galactic coordinates, the $V_{Macho}$ periods, 
the $V_{Macho}$ amplitudes, the $\sigma_V$, the chi-square of the phasing fit,
the mean V and R magnitudes, and the reddening-independent W magnitudes
(defined in Section 4).

\section{The Photometry}

The photometric calibration of the large MACHO database is challenging.
Only differential photometry is needed for a microlensing search, so individual
field zero-points
and a transformation to the
standard system are not priority tasks for the survey telescope.
The non-standard MACHO filters have been calibrated with standard 
Cousins photometry using Landolt standards transferred to
LMC fields. We have used these calibrations in the present work.  
The $V_{Macho}$ and $R_{Macho}$ magnitudes are transformed to standard 
Cousins V and R magnitudes using the following equations:
$$V = 23.67+1.0026~ V_M - 0.156~ (V_M - R_M)$$
$$R = 23.48+1.0044~ R_M + 0.182~ (V_M - R_M),$$
where the zero-point is adjusted to match the photometry of stars
in common with Cook (1987) and Walker \& Mack (1986) Baade's window (BW).

The 24 fields monitored have zero-points that may differ at the 
0.15 mag level at most.   Cross checks were made with the RRab found in 
the overlapping regions, and with the OGLE data (V and I photometry) in BW.
However, in order to avoid systematic effects, 
our analysis will be restricted to a differential approach, (i.e. 
determining relative distances between the bulge and Sgr).

In the regions where two or more fields overlap,
we identified a total of 44 bulge RRab with $0.4 < P < 0.7$, where
P is the pulsation period in days.
Each photometric measurement is flagged for
errors due to crowding, radiation 
blemishes, array defects, bad seeing, and edge effects.
A few of the variables were not found in both overlapping fields
(some rejected a priori for being period aliases, others 
was not considered as variables, having fewer unflagged points).
Using our internal redundancy, we estimate an upper limit of 93\%
for the completeness of our RRab sample within the period/amplitude
cuts selected.

Using the 44 bulge RR Lyrae stars in overlap regions,
we have evaluated the internal uncertainty
in the photometry ($\sigma_{V_M},~\sigma_{R_M}$), 
astrometry ($\sigma_{\alpha},~\sigma_{\delta}$), period ($\sigma_P$), and 
amplitude ($\sigma_{AV_M},~\sigma_{AR_M}$) determinations. 
We obtain the following errors:
$\sigma {V_M} = 0.120$,
$\sigma {R_M} = 0.114$,
$\sigma_{\alpha} = 0.58\arcsec$,
$\sigma_{\delta} = 0.37\arcsec$, 
$\sigma_P = 0.000054^d$, 
$\sigma_{A V_M} = 0.12$, and 
$\sigma_{A R_M} = 0.06$.

The OGLE data in BW (Udalski et al. 1994)
provide an independent check on the  errors and
completeness of our sample.  The 9 OGLE CCD's cover the same total area as 
MACHO field 119.
In the overlapping field with OGLE we find 44 RRab within the same period range.
Of these, 25 are in common, and the comparison yields:
$\Delta {\alpha}= 0.16"\pm 0.54\arcsec$,
$\Delta {\delta}= 0.24"\pm 0.42\arcsec$, and
$\Delta P = 0^d \pm 0.00005^d ,$
where $\Delta  = {Macho} - {Ogle}.$
These values are consistent with the internal estimates obtained 
from the overlapping regions of the MACHO fields.

The field of Alard (1996) also overlaps with some of our fields. In this
overlap region we find 11 RRab belonging to the Sgr dwarf, 
and 6 in common with Alard (1996).
The differences between the samples are not 
surprising given the different selection criteria adopted.
We emphasize that many of
the stars that do not pass our initial periodic variable cuts
will be recovered when the 2nd and 3rd year data are analyzed.

\section{Reddening}

Figure 1 shows the
color-magnitude diagram for all the RRab in the MACHO 1993
bulge sample, including RRab belonging
to the Sgr dwarf galaxy. The appearance of this color-magnitude diagram
is largely dominated by reddening. The direction of the
reddening vector is indicated.
The reddening is patchy in the MACHO fields towards the bulge, 
ranging from $E(B-V) = 0.2$ in the outer fields, to $E(B-V) \geq 2$ in the 
most obscured regions. We will then use
reddening-independent magnitudes that assume a standard extinction law
for our comparison.  These magnitudes, defined as 
$W_V =  V - 3.97 \times (V - R)$, are listed in the last column of Table 1. 
Note that in the most heavily reddened fields some of the faintest variables
will be lost.  Because we only reliably detect RRab which are brighter than
$V \sim 19.5$, the distance to which we can detect
RRab with $0.4 < P < 0.7$ and $A_V > 0.2$ mag depends on the reddening.
For $E(B-V) = 0.5$ (typical of the BW field), we would
detect RRab located 
well beyond the known distance of Sgr.

In order to avoid the most heavily reddened regions (where 
$A_V / E_{B-V}$ may be significantly different from 3.1), 
the relative distances will be measured using only RRab with $V-R < 0.82$, and
located in fields with $b < -4^{\circ}$.

\section{The Periods}

There is a strong dependence of RR ab period and luminosity
with [Fe/H] (e.g. Sandage
1993a, Jones et al. 1992), in the sense that the more metal--poor
RR ab tend to have longer periods. The effect of metallicity in the
period--amplitude plane is clearly illustrated by Figs. 10--12 of
Jones et al. (1992). There is also a period dependence on $T_{eff}$
-- or color -- (e.g. Sandage 1993b, Carney et al. 1992), which cannot be
investigated with the present data due to the effect of differential
reddening.

These P-L-Z relations have to be taken into account when determining relative
distances. Therefore, the relative
distance between the RRab in the bulge and in Sgr is computed here
using only the RRab in
the bulge covering the same period range as the Sgr RRab ($i.e.~
0.46^d < P < 0.66^d$). 

The bulge RRab have a mean $[Fe/H] =
-1.0 \pm 0.16$ (Walker \& Terndrup 1991).
The periods listed in Table 1 support
the conclusion that the Sgr RRab
have longer mean periods, and are therefore more metal-poor {\it in the
mean} than the Galactic RRab in these fields (Alard 1996). 
The period distribution in Sgr resembles that of the LMC, shown by
Alcock et al. (1996).  Even though Sgr also has a metal-rich component 
(Sarajedini \& Layden 1995), the bulk of the RR Lyrae stars in this galaxy
must be produced by the metal-poor population with $[Fe/H] = -1.3$, 
as argued by Mateo et al. (1995).
The probability of the formation of RR Lyrae stars in a metal-poor population
is about a factor of $50$ larger than in a metal-rich population (Suntzeff et
al. 1991).

\section{The Structure of Sgr}

The distribution on the sky of the RRab listed in Table 1
is shown in Figure 2.  Most of the Sgr RRab
are located in the MACHO fields which are well
off the Galactic minor axis, and in the lower latitude
fields, with only a few of them in the fields closer to the Galactic center, 
including BW. 
The MACHO fields are located in the northern--most extension of Sgr 
(in Galactic coordinates), reaching 3$^{\circ}$ further away from its center
than the fields studied by Alard (1996), who found a declining number of
RR Lyrae stars in that direction.

The outer contours of Sgr in the discovery paper by Ibata et al. (1994) cover
about $10^{\circ}$ along its major axis. However, this galaxy is much more
extended.
In retrospect, perhaps the earliest observational
record of Sgr is the excess of blue stars at about $V = 18$ in the luminosity
functions of Rodgers \& Harding (1989) for a bulge field at $l,~b =
10^{\circ},~-22.3^{\circ}$. These stars with $V \approx 18$, and
$0.3 < (B-V)_0 < 0.5$ would be horizontal branch stars at the distance of Sgr.
Rodgers et al. (1990) obtained radial velocities and calcium abundances for
18 stars within this color range. Two of these stars (\# 2730 and
\# 3844) have heliocentric radial velocities consistent with Sgr membership 
($V_{2730} = 162 \pm 38~ km ~s^{-1}$, and
$V_{3844} = 152 \pm 8~ km ~s^{-1}$). Stars associated with Sgr have been 
found in two other fields next to that of Rodgers \& Harding (1989):
Mateo et al. (1996) discovered three RRab associated with Sgr in a field at
$l,~b = 8.8^{\circ},~-23.3^{\circ}$, and
Fahlman et al. (1996) detected a sequence of Sgr stars in their 
color-magnitude diagram for a field at $l,~b = 9^{\circ},~-23^{\circ}$. 
The presence of Sgr RRab in the MACHO fields
implies that the major axis of Sgr is at least $20^{\circ}$ in size,
while the minor axis extension is at least $7^{\circ}$ in size.

Figure 3 shows the distribution of distance moduli for RRab
detected in the MACHO 1993 data. The highest peak at $W_V=14.6$ is due to 
the bulge RRab. The second peak at $W_V=16.8$
is real, and due to the Sgr members. 
Adopting $R_{\odot} = 8$ kpc (see the most recent determinations and discussion
by Carney et al. 1995), the difference between these
peaks, $\Delta m-M = 2.2 \pm 0.1$ mag, locates the Sgr dwarf at 
$D = 22 \pm 1$ kpc.

Previous distance estimates to Sgr are listed in Table 2. These
mean distance determinations range from 24.0 to 27.6 kpc. The present distance
D = 22 kpc marks the low end of this distribution. In particular,
it is significantly different than the distances of three RRab 
-- $D = 26.4, 27.4,$ and $28.2$ kpc -- in the opposite side of Sgr, measured by 
Mateo et al. (1996).
We argue that the distance spread is real, and due to the fact that
Sgr is inclined along the line of sight.

This distance
measurement is relative to the Galactic bulge, and largely independent
of the RRab absolute magnitude calibration. We have checked for a dependence
of this distance with metallicity, by dividing
the sample into metal--poor and metal--rich
RRab using the period--amplitude diagram. Even though the Sgr sample is small, 
both longer period (metal--poor) and shorter period (metal--rich)
RRab yield similar results.
Other effects such as a barred bulge or differential reddening would make
the distance shorter.

The various measured distances of Sgr projected in the Galactic x-z plane
are plotted in Figure 4.
The maximum difference in distances along the line of sight is found between 
the determination of Mateo et al. (1996) and us. 
These are two of the most separated fields of Sgr
studied so far, about $20^{\circ}$ apart. The difference in distance 
moduli between these two independent determinations is $\Delta (m-M) = 0.47
~mags$, larger than the estimated errors listed in Table 2. 
The distance of Mateo et al. (1996) is the mean of
only three RRab, but it is confirmed by deep color--magnitude
diagrams that reach the turn--off of Sgr (Fahlman et al. 1996, 
Mateo et al. 1996), and therefore cannot be in
gross error.  Fahlman et al. (1996) list a range of distances, from 26.3
to 28.9 kpc, depending on the parameters adopted for their main sequence
fit. 

Sgr appears very elongated in the plane of the sky, 
with an axial ratio 3/1 (Ibata
et al. 1995). Two configurations can give rise to such a projected shape:
a flat disk seen edge-on, or an intrinsic
cigar--shape. 
However, an edge-on disk is ruled out by our observations, 
since this geometry
would yield similar distances at both extremes of the galaxy. Thus,
Sgr has a cigar shape.  Assuming axial symmetry, 
the line of sight depth of Sgr should be $\sim 4$ kpc. This is consistent
with the observed $FWHM = 0.35 \pm 0.05$ of the
magnitude distribution of RR Lyrae stars variables (this work, Alard 1996), 
again ruling out an edge-on disk with line of sight depth $\sim 10$ kpc.

This elongated configuration
is the predicted effect of a close encounter with the Milky Way,
``as tidal effects string stars out along the orbit"
(Velazquez \& White 1995), and as ``the sole effect of tides,
as seen from our galaxy, is an elongation of the dSph in the orbital plane"
(Piatek \& Pryor 1995). Numerical simulations of Sgr's past history
show that when the galaxy being disrupted 
is near pericenter, its major axis is approximately perpendicular to the
direction of the Galactic center, i.e. parallel to its orbital motion. 
Taking into account the radial
velocity $V = +160$ \kms, 
assuming that the direction of the elongation traces Sgr's orbit, and
projecting this as a straight line (first-order approximation), we find
that pericenter occurred 
at a Galactocentric distance of $R \approx 13$ kpc, on the opposite side of 
the Galactic plane, at $l,~b \approx (0^{\circ}, 12^{\circ})$.  Then, Sgr is 
presently moving away from the disk, having crossed the Galactic plane 
at a Galactocentric distance of about 14 kpc (equivalent to 3.5 Galactic
scalelengths $h_r$).

With the measured radial velocity and projection angles of its tangential 
motion, the orbit of Sgr is determined. However,
further confirmation of the orbital direction will come with the measurement
of accurate proper motions. We can say that
the present analysis agrees with the orbit computed by
Velazquez \& White (1995), having transverse velocity of 255 \kms
directed away from the Galactic plane, with pericenter and apocenter of
10 and 52 kpc, respectively, and orbital period of 0.76 Gyr. 
Better agreement is found with the orbit \#1 of Johnston et al. (1995),
with 1.08 Gyr period, and pericenter and apocenter of 13.4 and 81.5, 
respectively. In this configuration, the globular clusters associated
with Sgr (Da Costa \& Armandroff 1995) would be leading the orbit.

However, preliminary results from a proper motion survey show that
Sgr is moving towards the Galactic plane (Irwin \& Wyse 1996, private 
communication). Assuming that the distances listed in Table 2 are correct,
this would indicate that our main assumption that
Sgr is elongated {\it along} its orbit may not be valid.

\section{Conclusions}

We have identified 30 RRab members of Sgr, 
located at the northern edge of this galaxy (in Galactic coordinates). 
Their positions, magnitudes, amplitudes and periods are listed in Table 1.
The mean estimated
distance to these stars is 22 kpc. This is significantly closer than previous
distances measured in the center and southern side of Sgr
(in Galactic coordinates), as summarized in Table 2. We conclude that 
Sgr is $\sim 10$ kpc long, it has a cigar shape, and it is
inclined along the line of sight, with its northern side  
(in Galactic coordinates) being closer (see Figure 4). 
This 
information allows us
to trace the orbit of Sgr and determine its previous history.
Sgr is moving away from the Galactic plane, having passed pericenter
which is located in the opposite side of the plane. Its predicted
orbit agrees with
previous numerical simulations, although the precise orbit needs
to be confirmed with accurate proper motions.

The present work is also another step towards the determination
of the line of sight distribution of 
sources for observed microlensing events. In the case of
the Sgr dwarf galaxy, its location at 22 kpc makes a very favourable
configuration for microlensing by bulge objects at 8 kpc, and
since the observed number of microlensing events is about 100,
some of these source stars should be in Sgr.
However, stars in Sgr cannot
explain the high optical depth determined from clump giant sources.

\acknowledgements
{
We are very grateful for the skilled support given to our project 
by the technical staff at the Mt. Stromlo Observatory.  
Work performed at LLNL is supported by the DOE under contract W7405-ENG-48.
Work performed by the Center for Particle Astrophysics on the UC campuses
is supported in part by the Office of Science and Technology Centers of
NSF under cooperative agreement AST-8809616.
Work performed at MSSSO is supported by the Bilateral Science 
and Technology Program of the Australian Department of Industry, Technology
and Regional Development. 
WJS is supported by a PPARC Advanced Fellowship. 
KG acknowledges support from DOE Outstanding Junior Investigator,
Alfred P. Sloan, and Cottrell awards. 
CWS thanks the Sloan, Packard and Seaver 
Foundations for their generous support.
}

\clearpage

\begin{deluxetable}{rrrrrrrrrrrrrlll}
\small
\footnotesize
\tablewidth{0pt}
\scriptsize
\tablecaption{Position, magnitudes, periods and amplitudes for Sgr RRab}
\tablehead{
\multicolumn{1}{c}{Field}& 
\multicolumn{1}{c}{Location}& 
\multicolumn{1}{c}{RA$_{2000}$}& 
\multicolumn{1}{c}{DEC$_{2000}$}& 
\multicolumn{1}{c}{$l$}& 
\multicolumn{1}{c}{$b$}& 
\multicolumn{1}{c}{$P_{V_M}$}& 
\multicolumn{1}{c}{$A_{V_M}$}& 
\multicolumn{1}{c}{$\sigma V_M$}& 
\multicolumn{1}{c}{$\chi$}& 
\multicolumn{1}{c}{$V$}& 
\multicolumn{1}{c}{$R$}& 
\multicolumn{1}{c}{$W_{V}$}}
\startdata
119&  770&  18 01 37.26&  -30 01 10.6& 0.83& -3.55&  0.5983& 1.038&  0.021& 37& 18.64& 18.01& 16.15\\
162&  991&  18 16 26.30&  -26 23 10.2& 5.60& -4.67&  0.5352& 0.573&  0.016& 19& 17.76& 17.37& 16.20\\
159&  352&  18 15 37.10&  -25 45 45.6& 6.06& -4.21&  0.4961& 0.819&  0.012& 52& 18.69& 18.07& 16.23\\
159&  407&  18 15 39.42&  -25 37 23.2& 6.19& -4.15&  0.5771& 0.727&  0.042& 12& 18.70& 18.10& 16.32\\
101&  713&  18 05 35.60&  -27 06 35.4& 3.80& -2.89&  0.5614& 0.781&  0.029&  7& 19.11& 18.43& 16.41\\
125&  489&  18 11 35.69&  -31 15 39.1& 0.78& -6.03&  0.4698& 1.213& 0.019& 154& 18.48& 17.99& 16.52\\
118&  847&  17 59 10.31&  -29 39 15.1& 0.89& -2.91&  0.4986& 0.839& 0.133&  14& 19.30& 18.60& 16.53\\
114& 1804&  18 03 58.90&  -29 13 17.8& 1.78& -3.61&  0.4824& 0.910& 0.027&  11& 19.10& 18.54& 16.53\\
104& 1156&  18 04 13.39&  -28 06 50.4& 2.77& -3.11&  0.5701& 0.746& 0.028&  10& 19.24& 18.57& 16.57\\
109&  557&  18 03 19.17&  -28 10 32.4& 2.62& -2.97&  0.5618& 0.558& 0.008&   8& 19.24& 18.57& 16.59\\
162&  878&  18 17 27.83&  -26 38 43.6& 5.48& -4.99&  0.5183& 0.693& 0.026&  19& 18.70& 18.19& 16.69\\
161&  822&  18 14 19.90&  -26 21 43.1& 5.40& -4.24&  0.4910& 0.891& 0.025&   8& 19.67& 18.93& 16.71\\
128& 1754&  18 08 08.28&  -29 11 53.5& 2.24& -4.39&  0.5750& 0.944& 0.065&  16& 19.17& 18.56& 16.74\\
103&   94&  18 13 57.11&  -27 14 06.3& 4.58& -4.58&  0.4698& 1.245& 0.025&  48& 18.87& 18.33& 16.76\\
125&  567&  18 13 32.26&  -31 10 37.4& 1.05& -6.36&  0.6343& 0.741& 0.012&  43& 18.52& 18.09& 16.81\\
105&  232&  18 07 39.45&  -27 53 13.1& 3.34& -3.66&  0.5887& 0.632& 0.033&   5& 19.66& 18.95& 16.83\\
167&  992&  18 14 10.55&  -27 08 20.1& 4.69& -4.58&  0.5630& 0.886& 0.039&  21& 19.16& 18.57& 16.83\\
124&  687&  18 07 17.53&  -31 24 56.6& 0.20& -5.29&  0.5733& 0.850& 0.013&  16& 19.05& 18.49& 16.84\\
162&  942&  18 17 20.76&  -26 25 26.3& 5.66& -4.86&  0.6046& 0.624& 0.026&   6& 19.48& 18.82& 16.87\\
115& 1265&  18 09 40.37&  -29 40 40.7& 1.98& -4.91&  0.5736& 0.901& 0.033&  16& 18.90& 18.39& 16.87\\
128& 2147&  18 08 12.33&  -28 58 03.8& 2.45& -4.29&  0.6434& 0.892& 0.047&   8& 19.63& 18.95& 16.93\\
125&  149&  18 10 17.27&  -30 38 16.2& 1.19& -5.49&  0.5712& 0.753& 0.016&  17& 19.26& 18.69& 16.99\\
116&  318&  18 11 56.42&  -29 34 29.0& 2.31& -5.30&  0.6256& 0.513& 0.006&  12& 18.87& 18.41& 17.05\\
114& 1463&  18 04 58.98&  -29 28 56.3& 1.66& -3.93&  0.6072& 0.422& 0.023&   4& 19.38& 18.79& 17.06\\
116&  547&  18 12 04.01&  -29 39 51.5& 2.24& -5.36&  0.6478& 0.796& 0.038&  13& 18.95& 18.48& 17.10\\
167& 1101&  18 13 54.75&  -26 49 29.4& 4.94& -4.38&  0.5663& 0.383& 0.014&  12& 18.43& 18.10& 17.12\\
162&  858&  18 17 47.03&  -26 38 57.9& 5.51& -5.06&  0.5418& 0.708& 0.067&  10& 19.12& 18.62& 17.14\\
111&  203&  18 12 09.70&  -28 40 11.3& 3.13& -4.91&  0.6459& 0.595& 0.054&   2& 19.82& 19.20& 17.34\\
125&  706&  18 13 20.51&  -31 03 52.9& 1.13& -6.27&  0.4891& 1.121& 0.035& 122& 18.80& 18.44& 17.38\\
162&  730&  18 15 23.28&  -26 34 53.0& 5.31& -4.55&  0.5247& 1.061& 0.064&  24& 19.54& 19.00& 17.41\\
\enddata
\end {deluxetable}

\begin{deluxetable}{lrllll}
\small
\footnotesize
\tablewidth{0pt}
\scriptsize
\tablecaption{Distance Estimates for Sgr }
\tablehead{
\multicolumn{1}{c}{$l$}&
\multicolumn{1}{c}{$b$}&
\multicolumn{1}{c}{$m-M$}&
\multicolumn{1}{c}{$D~(kpc)$}&
\multicolumn{1}{c}{Method}&
\multicolumn{1}{c}{Reference}}
\startdata
4.0& -4.0 &16.71 &$22.0\pm 1.0$ &RRab    &MACHO ($R_{bulge} = 8$ kpc)\nl
4.0& -8.0 &16.90 &$24.0\pm 2.0$ &RRab    &Alard 1996\nl
5.6&-14.1 &16.99 &$25.0$        &CMD     &Ibata et al. 1995\nl
5.6&-14.1 &17.00 &$25.1\pm 4.0$ &4 globulars&DaCosta \& Armandroff 1995\nl
5.6&-14.1 &17.02 &$25.4\pm 1.0$ &RHB-RGBC&Sarajedini \& Layden 1995\nl
6.6&-16.3 &17.02 &$25.4\pm 2.4$ &RRab    &Mateo et al. 1995\nl
8.8&-23.3 &17.18 &$27.3\pm 1.0$ &RRab, CMD&Mateo et al. 1996\nl
9.0&-23.0 &17.20 &$27.6\pm 1.3$ &CMD     &Fahlman et al. 1996\nl
\enddata
\end {deluxetable}

\clearpage

\begin{figure}
\plotone{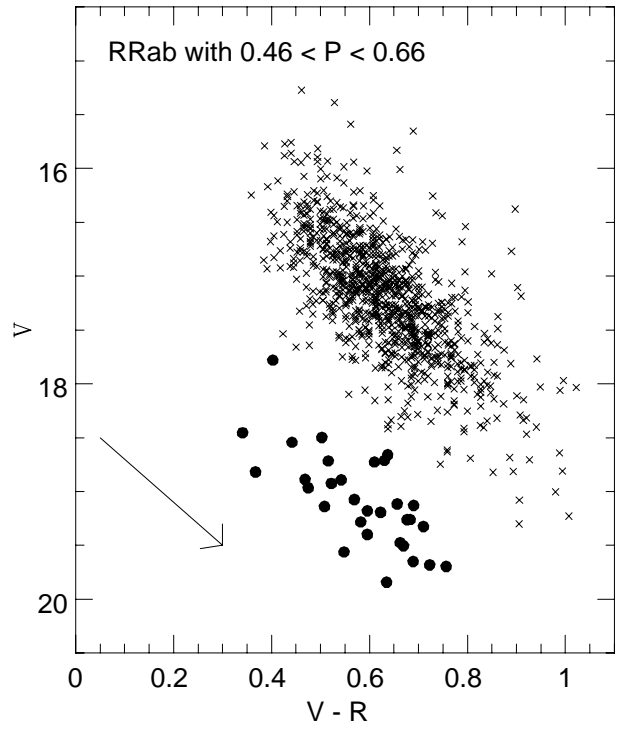}
\caption{ 
Color-magnitude diagram for the RRab in the MACHO 1993 
bulge sample. Note the group of fainter RRab belonging
to the Sgr dwarf galaxy (filled circles). The direction of the
reddening vector is indicated.
}
\end{figure}

\begin{figure}
\plotone{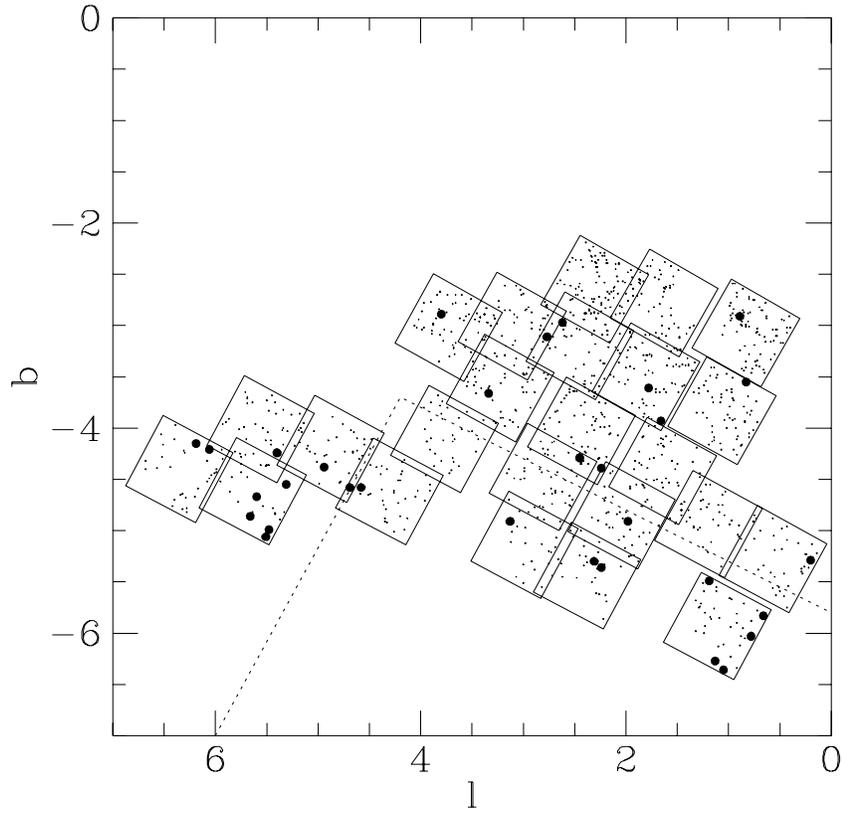}
\caption{ 
The 24 fields surveyed by MACHO in the first year of bulge
observations. The location of the bulge and Sgr RRab are indicated with 
dots and filled circles, respectively. The position of the field observed 
by Alard (1996) is marked with the dotted line.
}
\end{figure}

\begin{figure}
\plotone{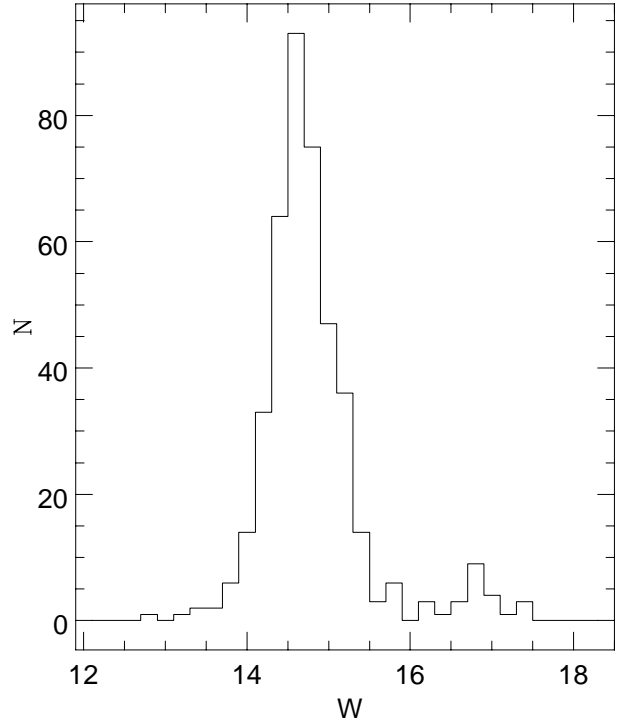}
\caption{ 
Magnitude distribution of the RRab with
$0.46 < P < 0.66$,  $b<-4^{\circ}$, and $V-R<0.8$
in the MACHO 1993 bulge data.
The peaks correspond to the Galactic bulge at $D = 8$ kpc, and
Sgr at $D = 22$ kpc.
}
\end{figure}

\begin{figure}
\plotone{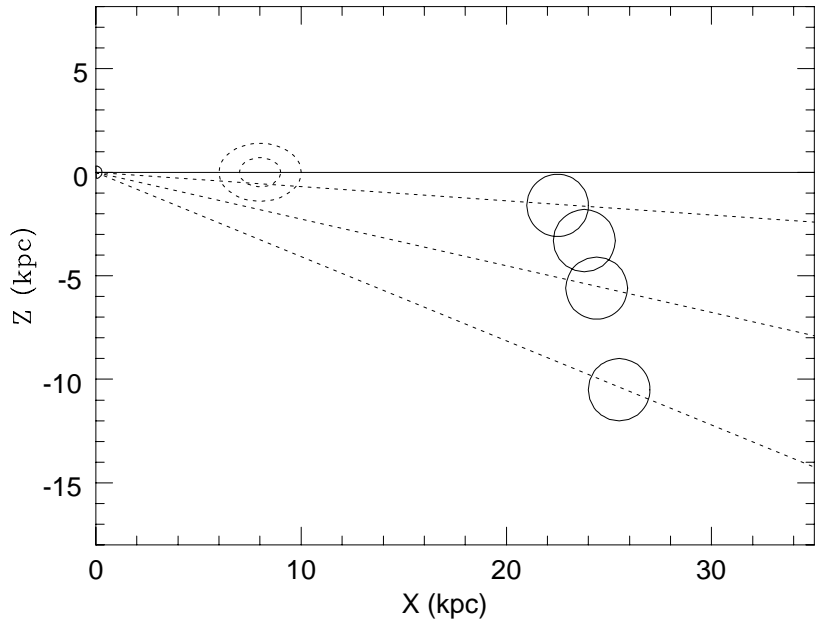}
\caption{ 
Measured distances of Sgr (big circles) projected in the Galactic x-z plane:
from top to bottom, the Sgr distance measured 
here, by Alard (1996), by Mateo et al. (1995), and by Mateo et al. (1996).
The Sun is located at $(0,~0)$, and the Galactic bulge at 
$(8,~0)$.
}
\end{figure}

\end{document}